%% file: paper.tex
\documentclass[11pt,a4paper]{article}
\usepackage[hyperref]{emnlp-ijcnlp-2019}
\aclfinalcopy
\usepackage{times}
\usepackage{latexsym}
\usepackage{url}
\usepackage{xargs}
\usepackage[colorinlistoftodos,prependcaption,textsize=tiny,color=lightgray]
{todonotes}
\newcommandx{\unsure}[2][1=]{\todo[#1]{#2}}
\newcommandx{\change}[2][1=]{\todo[#1]{#2}}
\newcommandx{\info}[2][1=]{\todo[#1]{#2}}
\newcommandx{\improvement}[2][1=]{\todo[#1]{#2}}
\newcommandx{\thiswillnotshow}[2][1=]{\todo[disable,#1]{#2}}

\usepackage{tikz}
\usetikzlibrary{shapes,arrows}
\usepackage{mathtools}
\usepackage[euler-hat-accent]{eulervm}
\usepackage{amsmath}
\usepackage{amssymb}
\usepackage{amsfonts}
\usepackage{bm}
\usepackage{caption,subcaption}

\DeclareMathOperator{\relu}{relu}
\let\cos\relax\DeclareMathOperator*{\cos}{cos}
\let\sin\relax\DeclareMathOperator*{\sin}{sin}
\let\softmax\relax\DeclareMathOperator{\softmax}{softmax}

\def\sectionautorefname~#1\null{\S#1\null}
\def\subsectionautorefname~#1\null{\S#1\null}
\def\equationautorefname~#1\null{(#1)\null}

\usepackage{array, multirow, bigdelim, makecell, booktabs}

\hypersetup{colorlinks,linkcolor=,urlcolor=blue}

\usepackage{amsthm}
\theoremstyle{definition}

\title{Learning Invariant Representations of Social Media Users}

\author{Nicholas Andrews and Marcus Bishop \\
Human Language Technology Center of Excellence\\
Johns Hopkins University\\
{\tt \{noa,marcus.bishop\}@jhu.edu}}

\begin{document}
\maketitle
\input{abstract}

\input{intro}
\input{dataRepresentation}

\input{batch}
\input{model}
\input{encoder}
\input{loss}
\input{datasets}

\input{experiments}
\input{relatedWork}
\input{conclusion}

\bibliographystyle{acl_natbib}
\bibliography{paper}

\end{document}

%% file: abstract.tex
\begin{abstract}
The evolution of social media users' behavior
over time complicates user-level comparison tasks
such as verification, classification, clustering, and ranking.
As a result, na\"ive approaches may fail to generalize to new users
or even to future observations of previously known users.
In this paper, we propose a novel procedure to learn
a mapping from short episodes of user activity on social
media to a vector space in which the distance between points
captures the similarity of the corresponding users' invariant features.
We fit the model by optimizing a surrogate metric learning
objective over a large corpus of unlabeled social media
content. Once learned, the mapping may be applied to users
not seen at training time and enables efficient comparisons
of users in the resulting vector space. We present a
comprehensive evaluation to validate the benefits of the
proposed approach using data from Reddit, Twitter, and Wikipedia.
\end{abstract}

%% file: intro.tex

\begin{figure*}
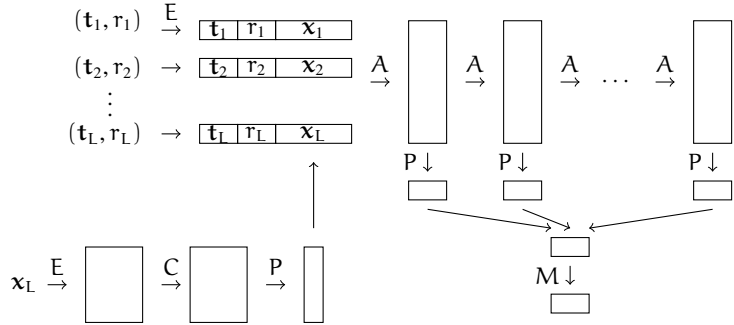

\begin{center}
\include{fig}
\end{center}
\caption{The map $f_{\bm{\theta}}$
takes an episode as input and outputs a vector.
Here $A$ denotes a multi-head self-attention layer,
$C$ a stack of 1D~convolutions,
$E$ an embedding lookup,
$M$ an MLP, and
$P$ a pooling layer.
}
\label{fig:model}
\end{figure*}

\section{Introduction}

Social media presents a number of challenges for characterizing user
behavior, chief among them that the topics of discussion and their participants
evolve over time. This makes it difficult to understand and combat
harmful behavior, such as election interference or
radicalization~\cite{thompson2011radicalization,mihaylov2016hunting,ferrara2016detection,keller2017manipulate}.

This work focuses on the fundamental
problem of learning to compare social media users. We
propose a procedure to learn embeddings of small samples of
users' online activity, which we call {\em episodes}.
This procedure involves learning the embedding using a
metric learning objective that causes episodes by the same
author to map to nearby points.
Through this embedding
users may be efficiently compared using cosine
similarity. This representation immediately enables several tasks:

\noindent{\bf Verification.} Determining if two episodes have the same author.\\
\noindent{\bf Classification.} Labeling authors via their
$k$-nearest neighbors.\\
\noindent{\bf Clustering.} Grouping users via off-the-shelf methods like $k$-means or
agglomerative clustering.\\
\noindent{\bf Ranking and retrieval.} Sorting episodes according to their
distances to a given episode.

The problem considered in this paper is most closely related to author
attribution on social media.
However, prior work in this area has primarily focused on
classifying an author as a member of a closed and typically small set of
authors~\cite{Stamatatos:2009:SMA:1527090.1527102,schwartz2013authorship,shrestha2017convolutional}.
In this paper, we are concerned with an \emph{open-world}
setting where we wish to characterize an {\em unbounded} number of users,
some observed at training time, some appearing only at test time.
A further challenge is that the episodes being compared may be drawn from different time
periods.
With these challenges in mind, the primary contributions described in this paper
are as follows:

\noindent{\bf \autoref{sec:batch}} A training strategy
in which a user's history is dynamically
sampled at training time to yield multiple short episodes drawn from different
time periods as a means of learning invariant features of the user's
identity;

\noindent{\bf \autoref{sec:model}} A user embedding
that can be trained end-to-end and which incorporates text, timing, and context
features from a sequence of posts;

\noindent{\bf \autoref{sec:datasets}} Reddit and Twitter benchmark
corpora for open-world author comparison tasks, which are
substantially larger than previously considered;

\noindent{\bf \autoref{sec:experiments}} Large-scale author ranking
and clustering experiments, as well as an application to Wikipedia
sockpuppet verification.

%% file: fig.tex
\begin{tikzpicture}[scale=0.125]

\draw (-5,13) node[anchor=east] {\small $\left(\bm{t}_1,r_1\right)$};
\draw [->] (-4,12) -- (-2,12) node[midway,above]{\footnotesize $E$};
\draw (0,11) rectangle (4,13);
\draw (2,12)  node {\small $\bm{t}_1$};
\draw (4,11) rectangle (8,13);
\draw (6,12)  node {\small $r_1$};
\draw (8,11) rectangle (16,13);
\draw (12,12)  node{\small $\bm{x}_1$};

\draw (-5,8) node[anchor=east] {\small $\left(\bm{t}_2,r_2\right)$};
\draw [->] (-4,8) -- (-2,8);
\draw (0,7) rectangle (4,9);
\draw (2,8)  node{\small $\bm{t}_2$};
\draw (4,7) rectangle (8,9);
\draw (6,8)  node{\small $r_2$};
\draw (8,7) rectangle (16,9);
\draw (12,8)  node{\small $\bm{x}_2$};

\draw (-8,5) node[anchor=east] {\footnotesize $\vdots$};

\draw (-5,1) node[anchor=east]  {\small $\left(\bm{t}_L,r_L\right)$};
\draw [->] (-4,1) -- (-2,1);
\draw (0,0) rectangle (4,2);
\draw (2,1)  node{\small $\bm{t}_L$};
\draw (4,0) rectangle (8,2);
\draw (6,1)  node{\small $r_L$};
\draw (8,0) rectangle (16,2);
\draw (12,1)  node{\small $\bm{x}_L$};

\draw [->] (18,6.5) -- (20,6.5) node[midway,above]{\footnotesize $A$};
\draw (22,0) rectangle (26,13);

\draw [->] (28,6.5) -- (30,6.5) node[midway,above]{\footnotesize $A$};
\draw (32,0) rectangle (36,13);

\draw [->] (38,6.5) -- (40,6.5) node[midway,above]{\footnotesize $A$};
\draw (44,6.5) node {\small $\cdots$};
\draw [->] (48,6.5) -- (50,6.5) node[midway,above]{\footnotesize $A$};
\draw (52,0) rectangle (56,13);

\draw [->] (24,-1) -- (24,-3) node[midway,left]{\footnotesize $P$};
\draw [->] (34,-1) -- (34,-3) node[midway,left]{\footnotesize $P$};
\draw [->] (54,-1) -- (54,-3) node[midway,left]{\footnotesize $P$};

\draw (22,-4) rectangle (26,-6);
\draw (32,-4) rectangle (36,-6);
\draw (52,-4) rectangle (56,-6);
\draw [->] (24,-7) -- (37,-9);
\draw [->] (34,-7) -- (39,-9);
\draw [->] (54,-7) -- (41,-9);

\draw (37,-10) rectangle (41,-12);
\draw [->] (39,-13) -- (39,-15) node[midway,left]{\footnotesize $M$};
\draw (37,-16) rectangle (41,-18);

\draw (-16,-15) node[anchor=east]{\small $\bm{x}_L$};
\draw [->] (-16,-15) -- (-14,-15) node[midway,above]{\footnotesize $E$};
\draw (-12,-19) rectangle (-6,-11);
\draw [->] (-4,-15) -- (-2,-15) node[midway,above]{\footnotesize $C$};
\draw (-1,-19) rectangle (5,-11);
\draw [->] (7,-15) -- (9,-15) node[midway,above]{\footnotesize $P$};
\draw (11,-19) rectangle (13,-11);
\draw [->] (12,-9) -- (12,-2);

\end{tikzpicture}

%% file: dataRepresentation.tex
\section{Preliminaries}\label{sec:data}

Broadly speaking, a corpus of social media data consists of the
{\em actions} of a number of users.  Each action consists of all
available information from a given platform detailing what exactly the
user did, which for purposes of this work we take to include: (1) a
timestamp recording when the action occurred, from which we extract a
tuple $\bm{t}$ of temporal features, (2) unstructured text content
$\bm{x}$ of the action, and (3) a categorical feature~$r$ specifying 
the context of the action.
Thus an action is a tuple of the form $\left(\bm{t},\bm{x},r\right)$.
This formulation admits all three platforms considered in this work
and therefore serves as a good starting point. However, incorporating
features specific to particular platforms, such as image, network,
and moderation features, might also provide useful signal.

In our experiments we use a data-driven sub-word
representation~\cite{kudo2018subword} of $\bm{x}$, which 
admits multilingual and non-linguistic content,
as well as misspellings and abbreviations, all of which
useful in characterizing authors.
We use a simple discrete time feature for $\bm{t}$, namely the hour of the day, 
although others might be helpful,
such as durations between successive actions.
In our Reddit experiments we take $r$ to be the subreddit to which
a comment was posted.
On Twitter we take $r$ to be a flag indicating whether
the post was a tweet or a retweet.

%% file: batch.tex
\section{Learning Invariant Representations}\label{sec:batch}
We organize the actions of each user into 
short sequences of chronologically ordered and ideally contiguous actions,
which we call {\em episodes}. This paper is concerned with devising a notion of
distance between episodes for which 
episodes by the same author are closer to one another than episodes by different authors.
Such a distance function must necessarily be constructed on the
basis of {\em past} social media data. But in the future, authors' behavior will evolve
and new authors will emerge.

We would like episodes by the same author to be nearby, irrespective
of when those episodes took place, possibly {\em future} to the creation of the distance function.
A given user will discuss different topics, cycle through various moods, develop
new interests, and so on, but distinctive features like uncommon word
usage, misspellings, or patterns of activity will persist for longer and
therefore provide useful signal for the distance function.

We would also like the distance to be meaningful when applied to episodes
by users who didn't exist when the distance function was created.
To this end, the features it considers must necessarily
generalize to new users. For example, common stylometric features
will be shared
by many users, including new users,
but their particular combination is distinctive of
particular users~\cite{orebaugh2009classification,layton2010authorship}.

Rather than heuristically defining such a distance function, for example, based on 
word overlap between the textual content of the episodes, we instead introduce a
parameterized embedding $f_{\bm{\theta}}$ shown in~\autoref{fig:model} that 
provides a vector representation of an episode.
Then the desired distance between episodes can be taken to be the distance
between the corresponding vectors.
We fit the embedding $f_{\bm{\theta}}$ using {\em metric learning}
to simultaneously decrease the distance between episodes by
the same user
and increase the distance between episodes by different
users~\cite{bromley1994signature,wang2014learning}.

But doing so requires
knowledge of the true author of an episode, something which is not generally available.
Therefore we take account names to be an approximation of latent authorship.
Of course, account names are not always a
reliable indicator of authorship on social media, as the
same individual may use multiple accounts, and multiple individuals
may use the same account.
As such, we expect a small amount of label noise in our data,
to which neural networks have proven robust in
several domains~\cite{krause2016unreasonable,rolnick2017deep}.




We fit $f_{\bm{\theta}}$ to a corpus of social media data using stochastic gradient descent on batches
of examples, where each example consists of an episode of a given length drawn
uniformly at random from the \emph{full history} of each user's actions.\footnote{Different metric learning methods will sample
  users in different ways, for example to ensure a given ratio of examples of
  the same class. In this work we simply sample users uniformly at random.}
By construction, a metric learning objective with this batching scheme will encourage the embedding of episodes drawn
from the same user's history to be close. In order to accomplish this, the model will
need to distinguish
between ephemeral and invariant features of a user.
The invariant features are those that enable the model to consistently distinguish
a given users' episodes from those of \emph{all other users}.


%% file: model.tex
\section{The Model}\label{sec:model}

\begin{figure*}
\centering
  \begin{subfigure}{.5\textwidth}
    \centering
    \includegraphics[scale=0.4]{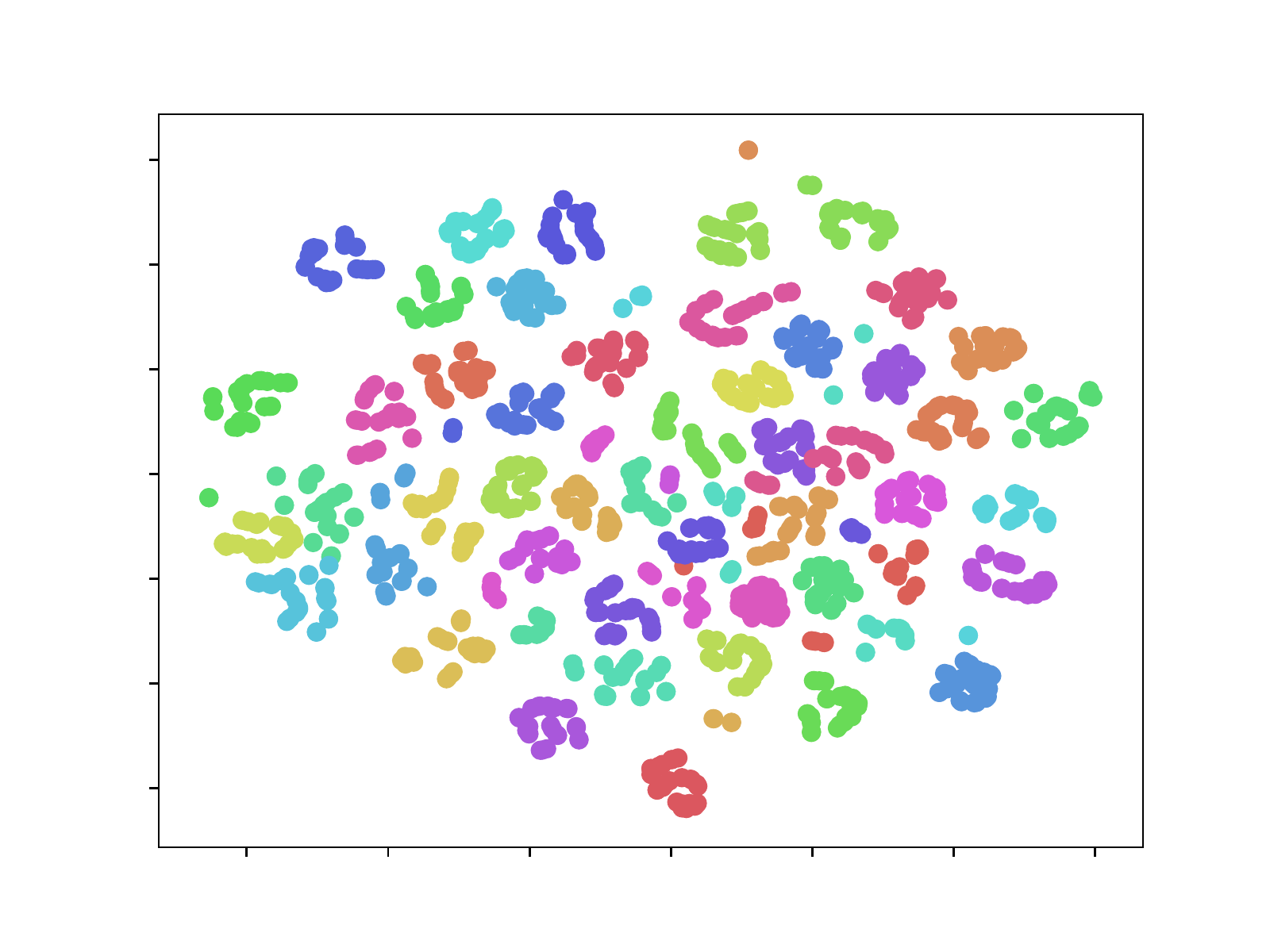}
    \caption{Embeddings obtained using {\bf SM} loss.}
  \end{subfigure}%
  \begin{subfigure}{.5\textwidth}
    \centering
    \includegraphics[scale=0.4]{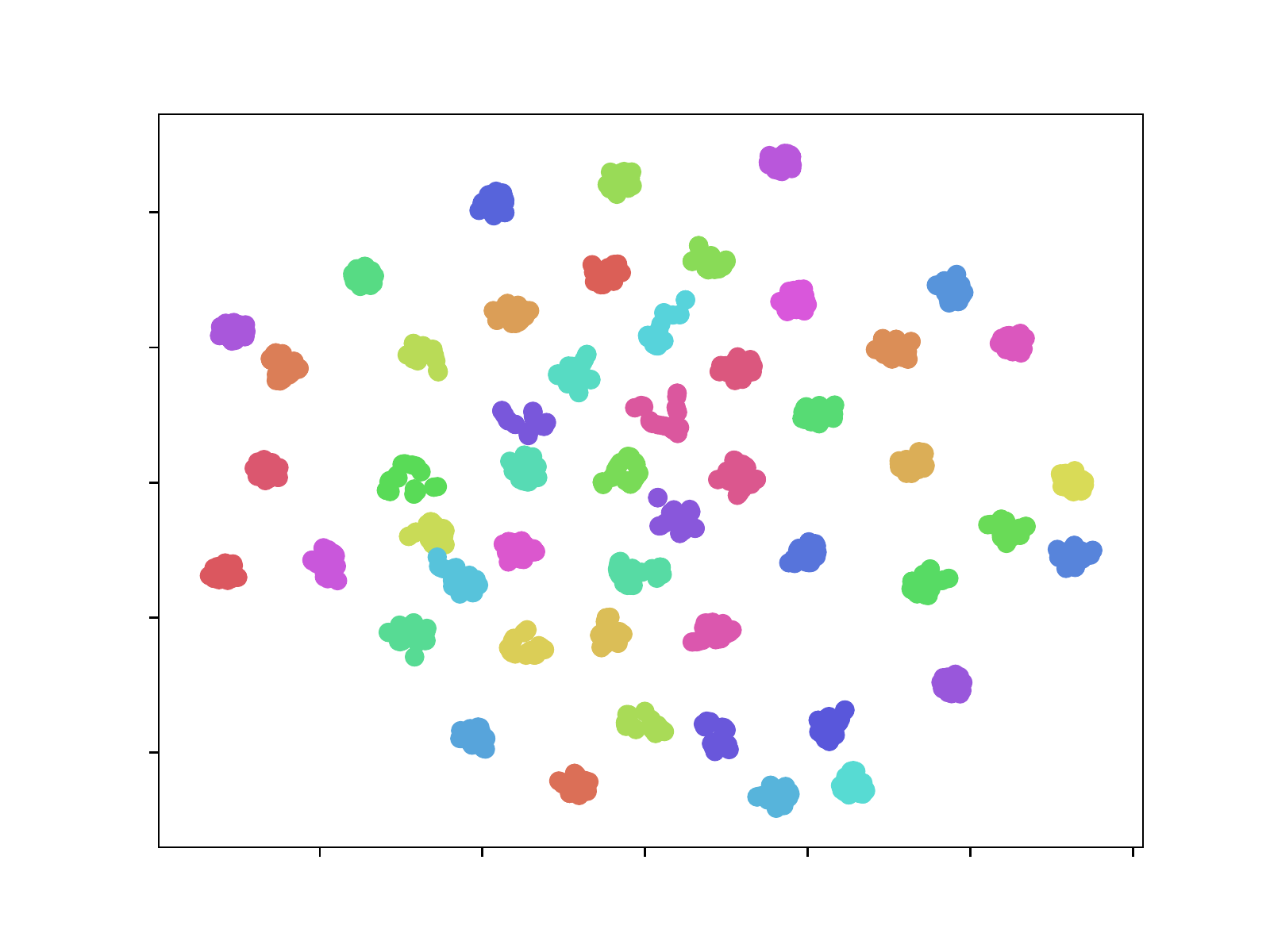}
    \caption{Embeddings obtained using {\bf AM} loss.}
    \label{fig:embeddingR}
  \end{subfigure}%
\caption{Projections of embeddings of user episodes. Each point is the result of mapping an episode to a single point in $\mathbb{R}^{512}$ and
projected to $\mathbb{R}^2$ using t-SNE.\ The colors of the points
correspond with the $50$ different authors of the underlying episodes. We emphasize that the episodes shown here were not seen by the model at training time.}
\label{fig:embedding}
\end{figure*}

We now describe a mapping $f_{\bm{\theta}}$
parameterized by a vector $\bm{\theta}$
from the space of user episodes
to $\mathbb{R}^D$.
The model is illustrated in \autoref{fig:model}.
This embedding induces a notion of distance between episodes
that depends on which of the two proposed
loss functions from \autoref{sec:loss} is used to train~$f_{\bm{\theta}}$.
We illustrate the embeddings resulting from both losses
in~\autoref{fig:embedding}.

%% file: encoder.tex
\subsection{The encoder}\label{sec:encoder}
One approach to define $f_{\bm{\theta}}$ might be
to manually define features of interest, such as
stylometric or surface features~\cite{sock2014lrec,sari2018topic}.
However, when large amounts
of data are available, it is preferable to use a data-driven approach
to representation learning. Therefore
we define $f_{\bm{\theta}}$ using a neural network as follows.
The network is illustrated in \autoref{fig:model}.

\vspace{5pt}
\noindent{\bf Encoding actions.}
First, we embed each action
$\left(\bm{t},\bm{x},r\right)$ of an episode.
We encode the time features $\bm{t}$ and the
context $r$, both assumed to be discrete,
using a learned embedding lookup.
We next embed every symbol of $\bm{x}$, again
using a learned embedding lookup,
and apply one-dimensional
convolutions of increasing widths over this list of vectors, 
similar to~\newcite{kim2014convolutional,shrestha2017convolutional}.  We then
apply the $\relu$ activation and
take the componentwise maximum of the list of vectors
to reduce the text content to a single, fixed-dimensional vector.
We optionally apply dropout at this stage if training.
Finally, we concatenate the time, text, and context vectors
to yield a single vector representing the action.

\vspace{5pt}
\noindent{\bf Embedding episodes.}
Next we combine the vector representations of the actions
of an episode.  For this purpose, one option is a recurrent neural
network (RNN).
However, recurrent models are biased due to processing inputs
sequentially, and suffer from vanishing and exploding gradients.
Therefore we propose the use of self-attention layers, which avoid
the sequential biases of RNNs and admit efficient implementations.

In our particular formulation, we use several layers of multi-head
self-attention, each taking 
the output of the previous layer as input;
architectural details of the encoder layers
follow those of the Transformer architecture proposed by~\newcite{vaswani2017attention}.
We apply mean pooling after every layer
to yield layer-specific embeddings,
which we concatenate. 
We project to the result to the
desired embedding dimension~$D$ using an MLP, both its input and output
batch normalized~\cite{ioffe2015batch}.

%% file: loss.tex
\subsection{The loss function}\label{sec:loss}
For the purpose of training the embedding $f_{\bm{\theta}}$
we compose it with a discriminative classifier
$g_{\bm{\phi}}:\mathbb{R}^D\to\mathbb{R}^Y$
with parameters $\bm{\phi}$
predicting the author of an episode,
where $Y$~is the number of authors in the training set.
We estimate $\bm{\theta}$ and $\bm{\phi}$ jointly using a standard
cross-entropy loss on a corpus of examples with their known authors.
Once the model is trained, the auxiliary projection $g_{\bm{\phi}}$
is discarded. Two possibilities for $g_{\bm{\phi}}$ are proposed below.

\vspace{5pt}

\noindent{\bf Softmax (SM).}
We introduce a weight matrix $\bm{W}\in\mathbb{R}^{Y\times D}$ and
define the map
$g_{\bm{\phi}}\left(\bm{z}\right)=\softmax \left({\bm{W}\bm{z}}\right)$
with parameters $\bm{\phi}=\bm{W}$.
When using this loss function, one compares
embeddings using Euclidean distance.

\vspace{5pt}

\noindent{\bf Angular margin  (AM).}
Following \newcite{deng2018arcface} we again
introduce a weight matrix $\bm{W}\in\mathbb{R}^{Y\times D}$
whose rows now serve as {\em class centers} for the training authors.
Given the embedding $\bm{z}\in\mathbb{R}^D$ of an episode,
let $\bm{z}'=\frac{\bm{z}}{\left\Vert\bm{z}\right\Vert}$ be the normalization
of $\bm{z}$ and let $\bm{W}'$ be obtained
from $\bm{W}$ by normalizing its rows. 
Then the entries of $\bm{w}=\bm{W}'\bm{z}'$ give the cosines of
the angles between $\bm{z}$ and the class centers.
Let $\bm{w}'$ be obtained from $\bm{w}$ by modifying the entry
corresponding with the correct author by
adding a fixed margin~$m>0$ to the corresponding angle.\footnote{
One way to calculate $\cos\left(\theta+m\right)$
from $\cos{\theta}$ is $\cos{\theta}\cos{m}-\sin{\theta}\sin{m}$
where $\sin{\theta}$ is calculated as
$\sqrt{1-\cos^2{\theta}}$. Note that this calculation
discards the sign of~$\theta$.}
Finally, define $g_{\bm{\phi}}\left(\bm{z}\right)
=\softmax\left(s\bm{w}'\right)$
where $s>0$ is a fixed {\em scale constant}.
When using this loss function, one compares embeddings using
cosine similarity.

%% file: datasets.tex
\section{Corpora for Large-Scale Author Identification}\label{sec:datasets}


\subsection{Reddit benchmark}

Reddit is a large, anonymous social media
platform with a permissive public
API. Using Reddit consists of reading and
posting {\em comments}, which consist of informal text,
primarily in English, each appearing within a
particular {\em subreddit}, which we treat as a categorical feature
providing useful contextual signal in characterizing users.


We introduce a new benchmark author identification corpus
derived from the API~\cite{gaffney2018caveat} containing
Reddit comments by 120,601 active users for training and
111,396 held-out users for evaluation. 
The training split contains posts published in
{\tt 2016-08} while the evaluation split contains posts
published in {\tt 2016-09}. In both cases, we restrict to users
publishing at least 100~comments but not more than~500. The lower
bound ensures that we have sufficient evidence for any given user for
training, while the upper bound is intended to mitigate the impact of bots
and atypical users.
The evaluation split is disjoint from the training split
and contains comments by 42,121 novel authors not contributing to the training
split.


\vspace{5pt}

\noindent {\bf Validation}.\ For model selection, we use the first
75\% of each user's chronologically ordered posts from the training set, with
the final 25\% reserved for validation. 
For example, in our ranking experiments described
in~\autoref{sec:experiments:reddit} we use these
held-out comments as candidate targets, using ranking performance to
inform hyper-parameter choice.






\subsection{Twitter benchmark}

The microblog domain is sufficiently distinct from Reddit that it is
suitable as an additional case study. For this purpose, we sample
169,663 active Twitter users from three months of 2016 as
separate training, development, and test sets (\texttt{2016-08}
through \texttt{2016-10}). We use three months because we rely on a
sub-sampled collection of Twitter, as little as 1\% of all posts published,
resulting in significantly fewer posts by each user than on Reddit.
Another consequence of this sub-sampling is that
the collection violates our assumptions regarding {\em contiguous}
user actions.

%% file: experiments.tex
\section{Experiments}\label{sec:experiments}
In the experiments described below, we refer to our method as IUR for
Invariant User Representations.

\subsection{Baseline methods}\label{sec:baselines}

In order to validate the merit of each of our
modeling contributions, we compare against three baseline
models described below.
To the best of our knowledge, we are the first to consider using
metric learning to learn embeddings from episodes of user activity.
We are also the first to consider doing so in open-world and large-scale 
settings. As such, the neural baseline described below uses the
training scheme proposed in this paper, and was further improved
to be more competitive with the proposed model.

\vspace{5pt}

\noindent \textbf{Neural author identification.}\
We use the architecture proposed by~\newcite{shrestha2017convolutional}
for closed-set author attribution
in place of our $f_{\bm{\theta}}$.  At the level of individual posts this
architecture is broadly similar to ours in that it applies 1D
convolutions to the text content. To extend it to episodes of
comments, we simply concatenate the text content
into a single sequence with a distinguished end-of-sequence marker. Note
that the timing and context features may also
be viewed as sequences, and in experiments with these features we run
a separate set of one-dimensional filters over them. All max-over-time
pooled features are concatenated depthwise. By itself, this model
failed to produce useful representations; we found it necessary to
apply the batch-normalized MLP described in \autoref{sec:encoder}
to the output layer before the loss.
To train the
model, we follow the procedure described in \autoref{sec:loss} to compose the embedding with the {\bf SM}
loss function, optimize the composition using cross-entropy loss, and
discard the {\bf SM} factor after training.

\vspace{5pt}

\noindent \textbf{Document vectors.}\ By concatenating all the textual
content of an episode we can view the episode as a single
document. This makes it straightforward to apply classical document
indexing methods to the resulting pseudo-document.
As a representative approach, we use TFIDF with cosine
distance~\cite{robertson2004understanding}.  We note that TFIDF is
also well-defined with respect to arbitrary bags-of-items, and we make
use of this fact to represent a user according to the sequence of
subreddits to which they post as a further
baseline in \autoref{sec:experiments:reddit}.

\vspace{5pt}

\noindent \textbf{Author verification models.}\ We use
the SCAP $n$-gram profile method of~\newcite{frantzeskou2007identifying}.
Two episodes are compared by calculating the size of the intersection of
their $n$-gram profiles. We use profiles of fixed length~64 in our
experiments.

\subsection{Model hyperparameters and training}\label{sec:training}
Below we list our hyperparameter choices for the IUR model, which
we define in \autoref{sec:model}.

For both Twitter and Reddit, we estimate the sub-word vocabulary on
training data using an inventory of 65,536~word pieces, including a
distinguished end-of-sequence symbol. We
truncate comments to 32~word pieces, padding if necessary.\footnote{In
  experiments not reported here, we have found that increasing the number of
  subwords per action increases performance but at the cost of slower training.}
We restrict to the 2048 most popular subreddits,
mapping all others to a distinguished {\tt unk} symbol.
We encode word pieces and subreddits as 256-long vectors.
The architecture for the text content uses four
convolutions of widths 2, 3, 4, 5 with 256 filters per convolution.
We use two layers of self-attention with 4
attention heads per layer, and hidden layers of size 512. Other
details such as use of layer normalization match the recommendations
of~\citet{vaswani2017attention}.

We train all variations
of the IUR for a fixed budget of 200,000
iterations of stochastic gradient descent with momentum 0.9 and a
piecewise linear learning rate schedule that starts at 0.1 and is
decreased by a factor of 10 at 100,000 and 150,000 iterations.
The final MLP has one hidden layer of dimension 512 with output
also of dimension $D=512$.
For the angular margin loss we take $m=0.5$ and $s=64$ as suggested in
\newcite{deng2018arcface}.

\subsection{Reddit ranking experiment}\label{sec:experiments:reddit}

\begin{table*}[h!]
  \begin{center}
    \footnotesize
    \tabcolsep=0.11cm
    \begin{tabular}{llcccccccc}
\toprule
\textbf{Input Features} & \textbf{Method} &\textbf{MRR} ($\Uparrow$)  &\textbf{MR} ($\Downarrow$) &\textbf{R@1} ($\Uparrow$) &\textbf{R@2}&\textbf{R@4}&\textbf{R@8}\\\midrule 
\multirow{5}{*}{text only}
&SCAP & 0.0057 &           31292 &          0.0035 &           0.004 &          0.0075 &          0.0085 \\ 
&TF-IDF (word)         & 0.071 &            5548 &           0.048 &           0.065 &           0.084 &            0.11  \\ 
&TF-IDF (char trigram) & 0.07 &            6264 &            0.05 &           0.066 &           0.081 &             0.1   \\ 
&\newcite{shrestha2017convolutional} & 0.081 &         660 &           0.052 &           0.071 &           0.094 &            0.12 \\ 
&IUR & {\bf 0.2} &      {\bf 88} &        {\bf  0.15} &       {\bf  0.19} &        {\bf  0.24} &        {\bf   0.29} \\ 
\midrule
\multirow{3}{*}{subreddit only}
&TF-IDF        &             0.1 &             305 &           0.068 &           0.091 &            0.12 &           0.16 \\ 
&\newcite{shrestha2017convolutional} & 0.18 &         110 &            0.12 &            0.16 &            0.21 &            0.26  \\ 
& IUR & {\bf 0.21} &     {\bf  64} &        {\bf  0.15} &         {\bf  0.2} &        {\bf  0.24} &             {\bf 0.3} \\  
\midrule
\multirow{5}{*}{text, subreddit, time}
&\newcite{shrestha2017convolutional}   &     0.39 &             8 &            0.31 &            0.38 &            0.45 &            0.51  \\ 
& IUR (softmax loss) & 0.38 &             9 &            0.31 &            0.38 &            0.44 &            0.49  \\  
& IUR (recurrent encoder) & 0.34 &         17 &            0.27 &            0.33 &            0.39 &            0.44 \\ 
& IUR (without time) & 0.48 &             3 &            0.41 &            0.48 &            0.55 &            0.61 \\ 
& IUR & {\bf 0.52} &            {\bf 2} &         {\bf 0.44} &       {\bf  0.52} &        {\bf  0.59} &           {\bf 0.65} \\ 
  \bottomrule
\end{tabular}
    \caption{Reddit author ranking results with 111,396 possible
      targets. The best results for each feature group are in
      printed in \textbf{bold}.
      The proposed Invariant User Representations are
      denoted IUR, with variations of the full model noted in
      parenthesis. MRR stands for the mean reciprocal rank, MR for
      median rank, and R@$k$ stands for recall at the top $k$ ranked
      episodes. Larger numbers are better ($\Uparrow$) except for MR where lower
      rank is better ($\Downarrow$). Metrics are computed over 25,000 queries.}
\label{tab:reddit}
\end{center}
\end{table*}

Given a query episode by a known user, our author ranking
experiment consists of returning a list of target episodes ranked
according to their similarity to the query.  The problem arises in the
moderation of social media content, when say, a user attempts to
circumvent an account ban by using another account.

\vspace{5pt}

\noindent \textbf{Experimental setup.}
Recall that we train all Reddit models on the {\tt 2016-08} split. 
In this experiment we draw episodes from the first
half of {\tt 2016-09} as {\em queries} and the second half of {\tt 2016-09}
as {\em targets}. Specifically, for each of 25,000 randomly selected users from
the {\tt 2016-09} split we randomly draw a query episode of
length~16 from among those posts published by that user before {\tt
2016-09-15}.  Then for each of the 111,396 users in the {\tt
2016-09} split we randomly draw a target episode of length~16
from among those posts published by that user on or after {\tt
2016-09-15}.  For each query, the goal of the experiment is to rank
the targets according to their likelihoods of being the {\em unique} 
target composed by the author of the query.

We compare models using mean reciprocal rank
(MRR), median rank (MR),
and recall-at-$k$ (R@$k$) for various $k$.
The MRR is the mean over all 25,000 queries
of the reciprocal of the position of
the correct target in the ranked list.
The MR is the median over the queries of the position of the correct target.
The R@$k$ is the proportion of the queries
for which the correct target appears among the first~$k$ ranked targets.

\vspace{5pt}

\noindent \textbf{Results.}\ The results of this experiment are shown
in~\autoref{tab:reddit}. For each combination of features considered,
the rankings based on the proposed IUR embeddings consistently
outperform all methods considered,
both neural and classical.
We also report results on several variations of our model, noted in parenthesis.
First, using the proposed architecture for $f_{\bm{\theta}}$ but the
softmax loss results in ranking performance comparable to the baseline system.
Second, using a recurrent architecture rather than self-attention to aggregate
information across an episode results in significantly worse performance.\footnote{We choose RNN hyper-parameters
such that the numbers of parameters of both models are on the same order of magnitude.}
Finally, omitting time features results in worse performance.

\vspace{5pt}

\noindent \textbf{Performance on novel users.}
As described above, the experiments presented in
\autoref{tab:reddit} involved ranking episodes by test authors,
some of whom had been seen during training, and some new to the model.
To better understand
the ability of the proposed embedding to generalize to new users, we
performed a further evaluation in which authors were restricted
to those {\em not} seen at training time. For the IUR incorporating all features,
this yielded a MRR of 0.50, while our extension
of~\newcite{shrestha2017convolutional} obtains 0.38 for the same
queries. Both methods produce salient embeddings of novel users, but
IUR retains an edge over the baseline.

\vspace{5pt}

\begin{figure}
  \centering
  \includegraphics[width=\linewidth]{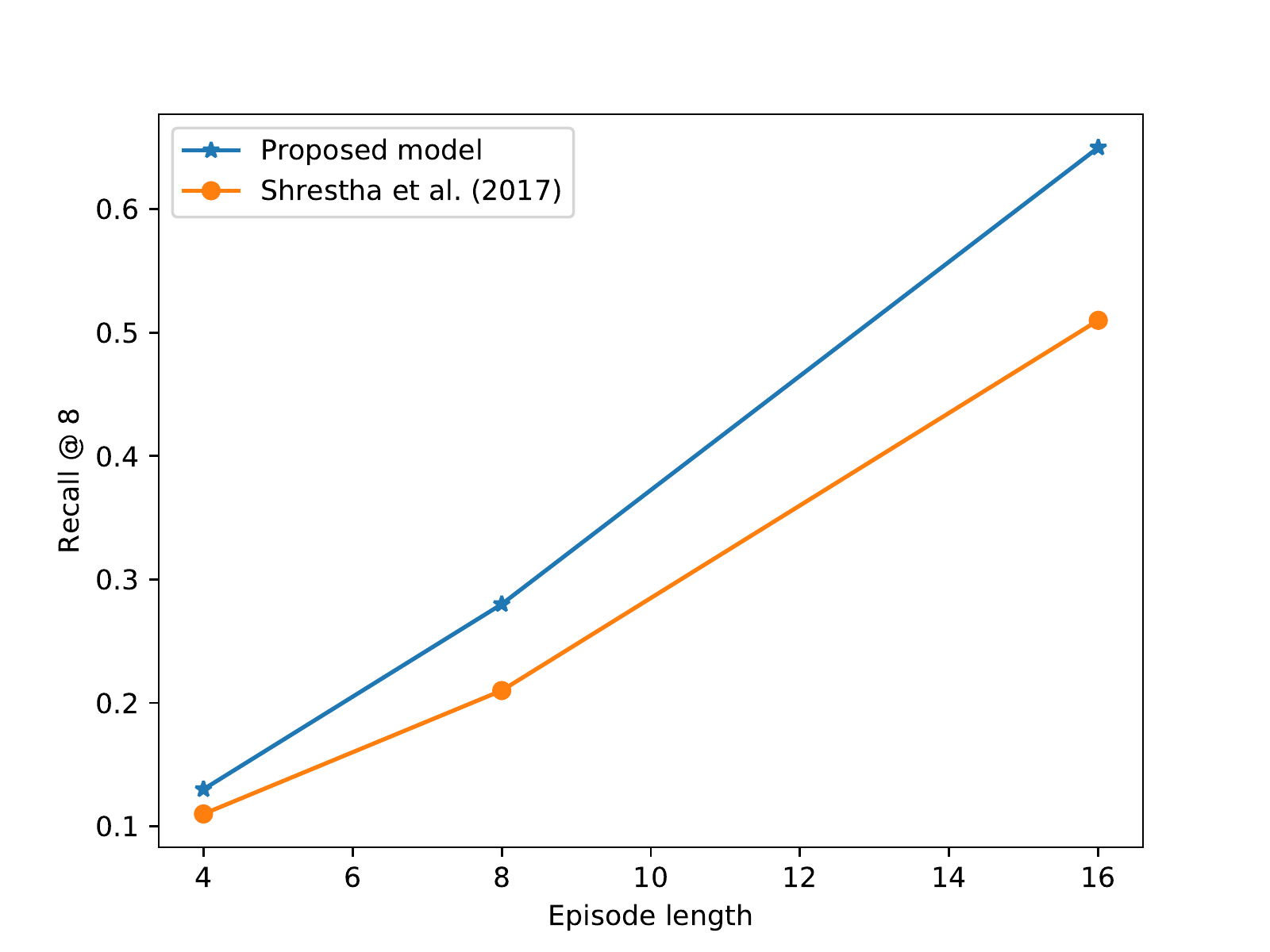}
  \caption{We report Recall@8 for different episode lengths using
    all features.}\label{fig:eplen}
\end{figure}

\noindent \textbf{Varying episode length.}
As described above, the experiments presented in
\autoref{tab:reddit} involved episodes of length exactly~16.
In~\autoref{fig:eplen}, we report results of a further ranking experiment in
which we vary the episode length, both at training time and at ranking time.
For both the proposed IUR and our extension
of~\newcite{shrestha2017convolutional},
performance increases as episode length increases.
Furthermore, even for the shortest episodes considered, the
proposed approach performs better.
This illustrates that the choice of episode length
should be decided on an application-specific basis. For example,
for social media moderation,
it may be desirable to quickly identify
problematic users on the basis of as few posts as possible.

\begin{table*}[h!]
  \begin{center}
    \footnotesize
    \tabcolsep=0.11cm
    \begin{tabular}{llrllllll}
        \toprule
\textbf{Method}&\textbf{MRR} ($\Uparrow$) &\textbf{MR} ($\Downarrow$) &\textbf{R@1} ($\Uparrow$) &\textbf{R@2}&\textbf{R@4}&\textbf{R@8}&\textbf{R@16}&\textbf{R@32} \\\midrule
TF-IDF (word)
&  0.060 &  4447 & 0.048 & 0.057 & 0.067 & 0.077 &  0.092 &   0.110\\
TF-IDF (char trigram)
&0.070 &  1622 & 0.052 &  0.064 & 0.078 &  0.095 &       0.120 &       0.140\\
\midrule
SCAP
 & 0.049 &  3582 & 0.037 &  0.044 &  0.053 &  0.065 &   0.08 &  0.098\\
\midrule
\citet{shrestha2017convolutional}
&0.056& 577 & 0.030 & 0.050 & 0.070 & 0.090 & 0.130 & 0.140\\
\midrule
IUR (text, time, context) &0.113&179&0.073&0.100&0.130&0.170&0.224&0.287\\
IUR (text only)&0.117&161&0.077&0.100&0.133&0.176&0.228&0.293\\
IUR (text, time) & {\bf 0.119} & {\bf 154} & {\bf 0.078} & {\bf 0.103} & {\bf 0.137} & {\bf 0.178} & {\bf 0.234} & {\bf 0.305} \\
\bottomrule
    \end{tabular}
    \caption{Twitter ranking results 
      with 25,000 queries and with 169,663 possible targets.}
    \label{tab:TwitterResults}
\end{center}
\end{table*}

\subsection{Twitter ranking experiment}\label{sec:exp:twitter}
We repeat the experiment described in~\autoref{sec:experiments:reddit}
using data from Twitter in place of Reddit, and with the further
difference that the queries were drawn from {\tt 2016-08} and the
targets from {\tt 2016-10} as a mitigation of Twitter's 1\%
censorship. Unlike the Reddit dataset, all three data splits contain
posts by exactly the same authors. The results are shown
in~\autoref{tab:TwitterResults}.

\subsection{Wikipedia sockpuppet verification}\label{sec:sock}

In this section we describe an experiment for the task of sockpuppet
verification on Wikipedia using the dataset collected
by~\newcite{sock2014lrec}. Wikipedia allows editors to open cases
against other editors for using suspected {\em sockpuppet} accounts
to promote their contributions. We have reorganized the dataset
into pairs of episodes by different accounts.
Half of our examples contain a pair deemed by the community
to have the same author, while half have been deemed
to have different authors.
The task is to predict whether a pair of episodes was composed
by the same author.

We are interested in whether the text-only version of
our IUR model, trained on Reddit data,
is able to transfer effectively to this task.
This domain is challenging because
in many cases sockpuppet accounts are trying to hide their identity,
and furthermore, Wikipedia talk pages contain domain-specific markup
which is difficult to reliably strip or normalize. Naturally we expect that
the identities of Wikipedia editors do not overlap with Reddit authors seen
at training time, since the data is drawn from different time periods and
from different platforms.

As a baseline, we compare to BERT, a generic text representation model trained
primarily on Wikipedia article text~\cite{devlin2018bert}. While BERT is
not specifically trained for author recognition tasks, BERT has obtained
state-of-the-art results in many pairwise text classification tasks including
natural language inference, question pair equivalence, question answering,
and paraphrase recognition. The BERT model used here has 110 million
parameters compared to 20~million for our embedding.


\noindent \textbf{Setup.}\ Because many comments are short, we pre-process the data to ensure that
each comment has at least 5~whitespace-separated tokens. We restrict
to users contributing at least 8~such comments.
This left us with 180~cases which we split into 72 for training, and
54 each for validation and testing.
We fine-tune both the cased and uncased pre-trained English BERT models
for our sockpuppet detection task using public models and
software.\footnote{\url{https://github.com/google-research/bert}}
In order to combine the comments comprising an episode for BERT,
we explored different strategies, including encoding
each comment separately. We found that simply combining
comments together and using a long sequence length of 512 gave the
best validation performance.
For our model, we
fine-tune by fitting an MLP on top of our embeddings using binary
cross entropy and keeping other parameters fixed.  Both methods are
tuned on validation data, and the best hyper-parameter configuration
is then evaluated on held-out test data.


\noindent \textbf{Results.}\ Results are reported
in~\autoref{tab:sock}. The best validation performance is obtained by
the cased BERT model. However, both BERT models appear to overfit the
training data as test performance is significantly lower. Regarding
the proposed IUR model, we see that its performance on validation data
is comparable to BERT while generalizing better to held-out test
data.
For reference, \newcite{solorio2013case}
report accuracy of 68.83 using the same data using a SVM with
hand-crafted features; however, neither their experimental splits nor
their model are available for purposes of a direct comparison.

\begin{table}[h!]
  \begin{center}
    \small
    \tabcolsep=0.11cm
    \begin{tabular}{lccc}
      \toprule
      & \textbf{Validation} & \textbf{Test}\\ \midrule
      Majority baseline & 0.5   & 0.5 \\
      BERT (uncased)    & 0.72  & 0.65 \\
      BERT (cased)      & {\bf 0.76}  & 0.61 \\
      IUR (text-only)   & 0.74  & {\bf 0.72} \\
      \bottomrule
    \end{tabular}
    \caption{Validation and test accuracy for the Wikipedia sockpuppet task. Best results in \textbf{bold}.}\label{tab:sock}
  \end{center}
\end{table}

\subsection{Clustering users}\label{sec:cluster}

For certain tasks it is useful to identify groups of accounts shared
by the same author or to identify groups of accounts behaving in a
similar fashion~\cite{solorio2013case,tsikerdekis2014multiple}.  To
this end, we experiment with how well a clustering algorithm can
partition authors on the basis of the cosine similarity of their IUR
episode embeddings.


\noindent \textbf{Procedure.}\ Using the pre-trained Reddit IUR model,
we embed five episodes of length 16 by 5000 users selected
uniformly at random, all drawn from the held-out \texttt{2016-09}
split. The embeddings are clustered using affinity propagation,
hiding both the identities of the users as well as the true number of
users from the algorithm~\cite{frey2007clustering}.
Ideally the algorithm will arrive at 5000 clusters, each containing
exactly five episodes by same author.
Clustering performance is evaluated using
mutual information (NMI), homogeneity (H), and completeness
(C)~\cite{rosenberg2007v}. NMI involves a ratio of the mutual
information of the clustering and ground truth. Homogeneity is a
measure of cluster purity. Completeness measures the extent to
which data points by the same author are elements of
the same cluster. All three measures lie in interval $\left[0, 1\right]$
where 1 is best. The results are shown in~\autoref{tab:clustering}.

\begin{table}[]
  \begin{center}
    \small
    \tabcolsep=0.11cm
    \begin{tabular}{lccc}
      \toprule
               & \textbf{NMI} & \textbf{H} & \textbf{C} \\ \midrule
      \newcite{shrestha2017convolutional} & 0.54 & 0.39 & 0.74                \\
      IUR           & $\mathbf{0.76}$       & {\bf 0.70}               & {\bf 0.84}                \\
      \midrule \\
    \end{tabular}
    \caption{Clustering performance on Reddit episodes using embeddings obtained with
      different methods.}
    \label{tab:clustering}
    \end{center}
  \end{table}

%% file: relatedWork.tex
\section{Related Work}\label{sec:lit}
\vspace{-3pt}


This work considers the problem of learning to compare users on social media.
A related task which has received considerably more attention
is predicting
user attributes~\cite{han2014text,sap2014developing,dredze2013carmen,culotta2015predicting,volkova2015inferring,goldin-etal-2018-native}.
The inferred user attributes have proven useful for social science and
public health research~\cite{mislove2011understanding,morgan2017predicting}.
While author attributes like gender or political leaning may be useful
for population-level studies, they are inadequate for identifying particular users.\footnote{We leave as future work the question of whether the episode
embeddings proposed in this paper are useful for attribute prediction.}


More generally, learning representations for downstream tasks using unsupervised training has recently emerged as an
effective way to mitigate the lack of task-specific training data~\cite{peters2018deep,devlin2018bert}.
 In the context of social media data, unsupervised methods have also been explored to obtain vector representations of
 individual posts on Twitter~\cite{dhingra2016tweet2vec,Vosoughi:2016:TLT:2911451.2914762}.
Our approach is distinguished from this prior work in several respects. First,
we embed episodes consisting of multiple documents, which involves aggregating features.
Second, for each document, we encode both textual features as well as associated meta-data. 
Finally, our training procedure is discriminative, embedding episodes
into a vector space with an immediately meaningful distance.


When social network structure is available, for example on Twitter
via {\em followers}, it may be used to learn 
user embeddings~\cite{tang2015line,Grover:2016:NSF:2939672.2939754,kipf2016semi}.
Graph representations have successfully been combined with
content-based features; for example, \newcite{benton2016learning} propose
matrix decomposition methods that exploit complementary features of
Twitter authors. Graph-based embeddings have proven useful in downstream
applications such as entity linking~\cite{yang2016toward}.
However, such methods are not applicable when network structure is
unavailable or
unreliable, such as with new users or on social media platforms like Reddit.
In this work, we are motivated in part by adversarial settings such as moderation, where
it is desirable to quickly identify the authorship
of novel users on the basis of sparse evidence.\footnote{
  Incorporating social network information in our model as additional
  features is in principle straightforward, requiring only minor architectural
  changes to the model; the metric learning procedure would otherwise remain the
  same.}


The most closely related work is author identification
on social media.
However, previous work in this area has largely
focused on distinguishing among small, closed sets of
authors rather than the open-world setting of this paper~\cite{mikros2013authorship,ge2016authorship}.
For example, \newcite{schwartz2013authorship} consider the problem of
assigning single tweets to one of a closed set of $1000$ authors.
\newcite{overdorf2016blogs} consider the problem of cross-domain
authorship attribution and consider $100$ users active on multiple
platforms.
In a different
direction, \newcite{sari2018topic} seek to identify stylistic features
contributing to successful author identification and consider
a closed set of $62$ authors.
In contrast, the present work
is concerned with problems involving several orders of magnitude more authors.
This scale precludes methods where similarity
between examples is expensive to compute, such as the method
of~\newcite{koppel2014determining}.


Prior work on detecting harmful behavior like hate speech has
focused on individual documents such as blog posts or
comments~\cite{spertus1997smokey,magu2017detecting,pavlopoulos2017deep,davidson2017automated,de2018modeling,basile2019semeval,zampieri2019semeval}. Recently, there
have been some efforts to incorporate user-level
information. For example, for the supervised task of abuse detection,
\newcite{mishra2018author} find consistent improvements from
incorporating user-level features.

%% file: conclusion.tex
\section{Conclusion}

Learning meaningful embeddings of social media users
on the basis of short episodes of activity poses a number of
challenges. This paper describes a novel approach to learning
such embeddings using metric learning coupled with a novel training regime
designed to learn invariant user representations. Our experiments
show that the proposed embeddings are robust with respect to both novel users
and data drawn from future time periods. To our knowledge, we are the
first to tackle open-world author ranking tasks by learning a vector
space with a meaningful distance.

There are several natural extensions of this work.
An immediate extension is to further scale
up the experiments to Web-scale datasets consisting of millions of users,
as has been successfully done for face recognition~\cite{kemelmacher2016megaface}.
Sorting episodes according to their
distances to a query can be made efficient using a number of
approximate nearest neighbor
techniques~\cite{indyk1998approximate,andoni2006near}.

We are also considering further applications of the proposed
approach beyond those in this paper.
For example, by restricting the features considered in the encoder
to text-alone or text and temporal features, it would be
interesting to explore cross-domain author attribution~\cite{stamatatos2018overview}.
It would also be interesting to explore community composition on the basis
of the proposed embeddings~\cite{newell2016user,Waller:2019:GSU:3308558.3313729}.

Finally, it bears mentioning that the proposed
model presents a double-edged sword:
methods designed to identify users engaging in harmful
behavior
could also be used to
identify authors with legitimate reasons to remain
anonymous, such as political dissidents, activists, or oppressed minorities.
On the other hand, methods similar to the proposed model could be developed
for such purposes and not shared with the broader community.
Therefore, as part of our effort to encourage positive applications, we release source code to reproduce our key results.\footnote{\url{http://github.com/noa/iur}.}

%% file: paper.bbl
\begin{thebibliography}{62}
\expandafter\ifx\csname natexlab\endcsname\relax\def\natexlab#1{#1}\fi

\bibitem[{Andoni and Indyk(2006)}]{andoni2006near}
Alexandr Andoni and Piotr Indyk. 2006.
\newblock Near-optimal hashing algorithms for approximate nearest neighbor in
  high dimensions.
\newblock In \emph{Proceedings of the 47th Annual IEEE Symposium on Foundations
  of Computer Science}, pages 459--468. IEEE Computer Society.

\bibitem[{Basile et~al.(2019)Basile, Bosco, Fersini, Nozza, Patti, Pardo,
  Rosso, and Sanguinetti}]{basile2019semeval}
Valerio Basile, Cristina Bosco, Elisabetta Fersini, Debora Nozza, Viviana
  Patti, Francisco Manuel~Rangel Pardo, Paolo Rosso, and Manuela Sanguinetti.
  2019.
\newblock Semeval-2019 task 5: Multilingual detection of hate speech against
  immigrants and women in twitter.
\newblock In \emph{Proceedings of the 13th International Workshop on Semantic
  Evaluation}, pages 54--63.

\bibitem[{Benton et~al.(2016)Benton, Arora, and Dredze}]{benton2016learning}
Adrian Benton, Raman Arora, and Mark Dredze. 2016.
\newblock Learning multiview embeddings of twitter users.
\newblock In \emph{Proceedings of the 54th Annual Meeting of the Association
  for Computational Linguistics (Volume 2: Short Papers)}, volume~2, pages
  14--19.

\bibitem[{Bromley et~al.(1994)Bromley, Guyon, LeCun, S{\"a}ckinger, and
  Shah}]{bromley1994signature}
Jane Bromley, Isabelle Guyon, Yann LeCun, Eduard S{\"a}ckinger, and Roopak
  Shah. 1994.
\newblock Signature verification using a ``siamese'' time delay neural network.
\newblock In \emph{Advances in neural information processing systems}, pages
  737--744.

\bibitem[{Culotta et~al.(2015)Culotta, Kumar, and
  Cutler}]{culotta2015predicting}
Aron Culotta, Nirmal~Ravi Kumar, and Jennifer Cutler. 2015.
\newblock Predicting the demographics of twitter users from website traffic
  data.
\newblock In \emph{AAAI}, pages 72--78.

\bibitem[{Davidson et~al.(2017)Davidson, Warmsley, Macy, and
  Weber}]{davidson2017automated}
Thomas Davidson, Dana Warmsley, Michael Macy, and Ingmar Weber. 2017.
\newblock Automated hate speech detection and the problem of offensive
  language.
\newblock In \emph{Eleventh International AAAI Conference on Web and Social
  Media}.

\bibitem[{Deng et~al.(2019)Deng, Guo, Xue, and Zafeiriou}]{deng2018arcface}
Jiankang Deng, Jia Guo, Niannan Xue, and Stefanos Zafeiriou. 2019.
\newblock Arcface: Additive angular margin loss for deep face recognition.
\newblock In \emph{Proceedings of the IEEE Conference on Computer Vision and
  Pattern Recognition}, pages 4690--4699.

\bibitem[{Devlin et~al.(2018)Devlin, Chang, Lee, and
  Toutanova}]{devlin2018bert}
Jacob Devlin, Ming-Wei Chang, Kenton Lee, and Kristina Toutanova. 2018.
\newblock Bert: Pre-training of deep bidirectional transformers for language
  understanding.
\newblock \emph{arXiv preprint arXiv:1810.04805}.

\bibitem[{Dhingra et~al.(2016)Dhingra, Zhou, Fitzpatrick, Muehl, and
  Cohen}]{dhingra2016tweet2vec}
Bhuwan Dhingra, Zhong Zhou, Dylan Fitzpatrick, Michael Muehl, and William
  Cohen. 2016.
\newblock Tweet2vec: Character-based distributed representations for social
  media.
\newblock In \emph{Proceedings of the 54th Annual Meeting of the Association
  for Computational Linguistics (Volume 2: Short Papers)}, pages 269--274.

\bibitem[{Dredze et~al.(2013)Dredze, Paul, Bergsma, and
  Tran}]{dredze2013carmen}
Mark Dredze, Michael~J Paul, Shane Bergsma, and Hieu Tran. 2013.
\newblock Carmen: A twitter geolocation system with applications to public
  health.
\newblock In \emph{Workshops at the Twenty-Seventh AAAI Conference on
  Artificial Intelligence}.

\bibitem[{Ferrara et~al.(2016)Ferrara, Varol, Menczer, and
  Flammini}]{ferrara2016detection}
Emilio Ferrara, Onur Varol, Filippo Menczer, and Alessandro Flammini. 2016.
\newblock Detection of promoted social media campaigns.
\newblock In \emph{tenth international AAAI conference on web and social
  media}.

\bibitem[{Frantzeskou et~al.(2007)Frantzeskou, Stamatatos, Gritzalis, Chaski,
  and Howald}]{frantzeskou2007identifying}
Georgia Frantzeskou, Efstathios Stamatatos, Stefanos Gritzalis, Carole~E
  Chaski, and Blake~Stephen Howald. 2007.
\newblock Identifying authorship by byte-level n-grams: The source code author
  profile (scap) method.
\newblock \emph{International Journal of Digital Evidence}, 6(1):1--18.

\bibitem[{Frey and Dueck(2007)}]{frey2007clustering}
Brendan~J Frey and Delbert Dueck. 2007.
\newblock Clustering by passing messages between data points.
\newblock \emph{science}, 315(5814):972--976.

\bibitem[{Gaffney and Matias(2018)}]{gaffney2018caveat}
Devin Gaffney and J~Nathan Matias. 2018.
\newblock Caveat emptor, computational social science: Large-scale missing data
  in a widely-published reddit corpus.
\newblock \emph{PloS one}, 13(7):e0200162.

\bibitem[{Ge et~al.(2016)Ge, Sun, and Smith}]{ge2016authorship}
Zhenhao Ge, Yufang Sun, and Mark~JT Smith. 2016.
\newblock Authorship attribution using a neural network language model.
\newblock In \emph{Thirtieth AAAI Conference on Artificial Intelligence}.

\bibitem[{Goldin et~al.(2018)Goldin, Rabinovich, and
  Wintner}]{goldin-etal-2018-native}
Gili Goldin, Ella Rabinovich, and Shuly Wintner. 2018.
\newblock Native language identification with user generated content.
\newblock In \emph{Proceedings of the 2018 Conference on Empirical Methods in
  Natural Language Processing}, pages 3591--3601, Brussels, Belgium.
  Association for Computational Linguistics.

\bibitem[{Grover and Leskovec(2016)}]{Grover:2016:NSF:2939672.2939754}
Aditya Grover and Jure Leskovec. 2016.
\newblock Node2vec: Scalable feature learning for networks.
\newblock In \emph{Proceedings of the 22Nd ACM SIGKDD International Conference
  on Knowledge Discovery and Data Mining}, KDD '16, pages 855--864, New York,
  NY, USA. ACM.

\bibitem[{Han et~al.(2014)Han, Cook, and Baldwin}]{han2014text}
Bo~Han, Paul Cook, and Timothy Baldwin. 2014.
\newblock Text-based twitter user geolocation prediction.
\newblock \emph{Journal of Artificial Intelligence Research}, 49:451--500.

\bibitem[{Indyk and Motwani(1998)}]{indyk1998approximate}
Piotr Indyk and Rajeev Motwani. 1998.
\newblock Approximate nearest neighbors: towards removing the curse of
  dimensionality.
\newblock In \emph{Proceedings of the thirtieth annual ACM symposium on Theory
  of computing}, pages 604--613. ACM.

\bibitem[{Ioffe and Szegedy(2015)}]{ioffe2015batch}
Sergey Ioffe and Christian Szegedy. 2015.
\newblock Batch normalization: Accelerating deep network training by reducing
  internal covariate shift.
\newblock \emph{arXiv preprint arXiv:1502.03167}.

\bibitem[{Keller et~al.(2017)Keller, Schoch, Stier, and
  Yang}]{keller2017manipulate}
Franziska~B Keller, David Schoch, Sebastian Stier, and JungHwan Yang. 2017.
\newblock How to manipulate social media: Analyzing political astroturfing
  using ground truth data from south korea.
\newblock In \emph{Eleventh International AAAI Conference on Web and Social
  Media}.

\bibitem[{Kemelmacher-Shlizerman et~al.(2016)Kemelmacher-Shlizerman, Seitz,
  Miller, and Brossard}]{kemelmacher2016megaface}
Ira Kemelmacher-Shlizerman, Steven~M Seitz, Daniel Miller, and Evan Brossard.
  2016.
\newblock The megaface benchmark: 1 million faces for recognition at scale.
\newblock In \emph{Proceedings of the IEEE Conference on Computer Vision and
  Pattern Recognition}, pages 4873--4882.

\bibitem[{Kim(2014)}]{kim2014convolutional}
Yoon Kim. 2014.
\newblock Convolutional neural networks for sentence classification.
\newblock \emph{arXiv preprint arXiv:1408.5882}.

\bibitem[{Kipf and Welling(2016)}]{kipf2016semi}
Thomas~N Kipf and Max Welling. 2016.
\newblock Semi-supervised classification with graph convolutional networks.
\newblock \emph{arXiv preprint arXiv:1609.02907}.

\bibitem[{Koppel and Winter(2014)}]{koppel2014determining}
Moshe Koppel and Yaron Winter. 2014.
\newblock Determining if two documents are written by the same author.
\newblock \emph{Journal of the Association for Information Science and
  Technology}, 65(1):178--187.

\bibitem[{Krause et~al.(2016)Krause, Sapp, Howard, Zhou, Toshev, Duerig,
  Philbin, and Fei-Fei}]{krause2016unreasonable}
Jonathan Krause, Benjamin Sapp, Andrew Howard, Howard Zhou, Alexander Toshev,
  Tom Duerig, James Philbin, and Li~Fei-Fei. 2016.
\newblock The unreasonable effectiveness of noisy data for fine-grained
  recognition.
\newblock In \emph{European Conference on Computer Vision}, pages 301--320.
  Springer.

\bibitem[{Kudo(2018)}]{kudo2018subword}
Taku Kudo. 2018.
\newblock Subword regularization: Improving neural network translation models
  with multiple subword candidates.
\newblock \emph{arXiv preprint arXiv:1804.10959}.

\bibitem[{Layton et~al.(2010)Layton, Watters, and
  Dazeley}]{layton2010authorship}
Robert Layton, Paul Watters, and Richard Dazeley. 2010.
\newblock Authorship attribution for twitter in 140 characters or less.
\newblock In \emph{2010 Second Cybercrime and Trustworthy Computing Workshop},
  pages 1--8. IEEE.

\bibitem[{Magu et~al.(2017)Magu, Joshi, and Luo}]{magu2017detecting}
Rijul Magu, Kshitij Joshi, and Jiebo Luo. 2017.
\newblock Detecting the hate code on social media.
\newblock In \emph{Eleventh International AAAI Conference on Web and Social
  Media}.

\bibitem[{Mihaylov and Nakov(2016)}]{mihaylov2016hunting}
Todor Mihaylov and Preslav Nakov. 2016.
\newblock Hunting for troll comments in news community forums.
\newblock In \emph{Proceedings of the 54th Annual Meeting of the Association
  for Computational Linguistics (Volume 2: Short Papers)}, pages 399--405.

\bibitem[{Mikros and Perifanos(2013)}]{mikros2013authorship}
George~K Mikros and Kostas Perifanos. 2013.
\newblock Authorship attribution in greek tweets using author's multilevel
  n-gram profiles.
\newblock In \emph{2013 AAAI Spring Symposium Series}.

\bibitem[{Mishra et~al.(2018)Mishra, Del~Tredici, Yannakoudakis, and
  Shutova}]{mishra2018author}
Pushkar Mishra, Marco Del~Tredici, Helen Yannakoudakis, and Ekaterina Shutova.
  2018.
\newblock Author profiling for abuse detection.
\newblock In \emph{Proceedings of the 27th International Conference on
  Computational Linguistics}, pages 1088--1098.

\bibitem[{Mislove et~al.(2011)Mislove, Lehmann, Ahn, Onnela, and
  Rosenquist}]{mislove2011understanding}
Alan Mislove, Sune Lehmann, Yong-Yeol Ahn, Jukka-Pekka Onnela, and J~Niels
  Rosenquist. 2011.
\newblock Understanding the demographics of twitter users.
\newblock In \emph{Fifth international AAAI conference on weblogs and social
  media}.

\bibitem[{Morgan-Lopez et~al.(2017)Morgan-Lopez, Kim, Chew, and
  Ruddle}]{morgan2017predicting}
Antonio~A Morgan-Lopez, Annice~E Kim, Robert~F Chew, and Paul Ruddle. 2017.
\newblock Predicting age groups of twitter users based on language and metadata
  features.
\newblock \emph{PloS one}, 12(8):e0183537.

\bibitem[{Newell et~al.(2016)Newell, Jurgens, Saleem, Vala, Sassine, Armstrong,
  and Ruths}]{newell2016user}
Edward Newell, David Jurgens, Haji~Mohammad Saleem, Hardik Vala, Jad Sassine,
  Caitrin Armstrong, and Derek Ruths. 2016.
\newblock User migration in online social networks: A case study on reddit
  during a period of community unrest.
\newblock In \emph{Tenth International AAAI Conference on Web and Social
  Media}.

\bibitem[{Orebaugh and Allnutt(2009)}]{orebaugh2009classification}
Angela Orebaugh and Jeremy Allnutt. 2009.
\newblock Classification of instant messaging communications for forensics
  analysis.
\newblock \emph{The International Journal of Forensic Computer Science},
  1:22--28.

\bibitem[{Overdorf and Greenstadt(2016)}]{overdorf2016blogs}
Rebekah Overdorf and Rachel Greenstadt. 2016.
\newblock Blogs, twitter feeds, and reddit comments: Cross-domain authorship
  attribution.
\newblock \emph{Proceedings on Privacy Enhancing Technologies},
  2016(3):155--171.

\bibitem[{Pavlopoulos et~al.(2017)Pavlopoulos, Malakasiotis, and
  Androutsopoulos}]{pavlopoulos2017deep}
John Pavlopoulos, Prodromos Malakasiotis, and Ion Androutsopoulos. 2017.
\newblock Deep learning for user comment moderation.
\newblock \emph{arXiv preprint arXiv:1705.09993}.

\bibitem[{Peters et~al.(2018)Peters, Neumann, Iyyer, Gardner, Clark, Lee, and
  Zettlemoyer}]{peters2018deep}
Matthew~E Peters, Mark Neumann, Mohit Iyyer, Matt Gardner, Christopher Clark,
  Kenton Lee, and Luke Zettlemoyer. 2018.
\newblock Deep contextualized word representations.
\newblock In \emph{Proceedings of NAACL-HLT}, pages 2227--2237.

\bibitem[{Robertson(2004)}]{robertson2004understanding}
Stephen Robertson. 2004.
\newblock Understanding inverse document frequency: on theoretical arguments
  for idf.
\newblock \emph{Journal of documentation}, 60(5):503--520.

\bibitem[{Rolnick et~al.(2017)Rolnick, Veit, Belongie, and
  Shavit}]{rolnick2017deep}
David Rolnick, Andreas Veit, Serge Belongie, and Nir Shavit. 2017.
\newblock Deep learning is robust to massive label noise.
\newblock \emph{arXiv preprint arXiv:1705.10694}.

\bibitem[{Rosenberg and Hirschberg(2007)}]{rosenberg2007v}
Andrew Rosenberg and Julia Hirschberg. 2007.
\newblock V-measure: A conditional entropy-based external cluster evaluation
  measure.
\newblock In \emph{Proceedings of the 2007 joint conference on empirical
  methods in natural language processing and computational natural language
  learning (EMNLP-CoNLL)}.

\bibitem[{Sap et~al.(2014)Sap, Park, Eichstaedt, Kern, Stillwell, Kosinski,
  Ungar, and Schwartz}]{sap2014developing}
Maarten Sap, Gregory Park, Johannes Eichstaedt, Margaret Kern, David Stillwell,
  Michal Kosinski, Lyle Ungar, and Hansen~Andrew Schwartz. 2014.
\newblock Developing age and gender predictive lexica over social media.
\newblock In \emph{Proceedings of the 2014 Conference on Empirical Methods in
  Natural Language Processing (EMNLP)}, pages 1146--1151.

\bibitem[{Sari et~al.(2018)Sari, Stevenson, and Vlachos}]{sari2018topic}
Yunita Sari, Mark Stevenson, and Andreas Vlachos. 2018.
\newblock Topic or style? exploring the most useful features for authorship
  attribution.
\newblock In \emph{Proceedings of the 27th International Conference on
  Computational Linguistics}, pages 343--353.

\bibitem[{Schwartz et~al.(2013)Schwartz, Tsur, Rappoport, and
  Koppel}]{schwartz2013authorship}
Roy Schwartz, Oren Tsur, Ari Rappoport, and Moshe Koppel. 2013.
\newblock Authorship attribution of micro-messages.
\newblock In \emph{Proceedings of the 2013 Conference on Empirical Methods in
  Natural Language Processing}, pages 1880--1891.

\bibitem[{Shrestha et~al.(2017)Shrestha, Sierra, Gonzalez, Montes, Rosso, and
  Solorio}]{shrestha2017convolutional}
Prasha Shrestha, Sebastian Sierra, Fabio Gonzalez, Manuel Montes, Paolo Rosso,
  and Thamar Solorio. 2017.
\newblock Convolutional neural networks for authorship attribution of short
  texts.
\newblock In \emph{Proceedings of the 15th Conference of the European Chapter
  of the Association for Computational Linguistics: Volume 2, Short Papers},
  volume~2, pages 669--674.

\bibitem[{Solorio et~al.(2013)Solorio, Hasan, and Mizan}]{solorio2013case}
Thamar Solorio, Ragib Hasan, and Mainul Mizan. 2013.
\newblock A case study of sockpuppet detection in wikipedia.
\newblock In \emph{Proceedings of the Workshop on Language Analysis in Social
  Media}, pages 59--68.

\bibitem[{Solorio et~al.(2014)Solorio, Hasan, and Mizan}]{sock2014lrec}
Thamar Solorio, Ragib Hasan, and Mainul Mizan. 2014.
\newblock Sockpuppet detection in wikipedia: A corpus of real-world deceptive
  writing for linking identities.
\newblock In \emph{Proceedings of the Ninth International Conference on
  Language Resources and Evaluation (LREC'14)}, Reykjavik, Iceland. European
  Language Resources Association (ELRA).

\bibitem[{Spertus(1997)}]{spertus1997smokey}
Ellen Spertus. 1997.
\newblock Smokey: Automatic recognition of hostile messages.
\newblock In \emph{AAAI/IAAI}, pages 1058--1065.

\bibitem[{Stamatatos(2009)}]{Stamatatos:2009:SMA:1527090.1527102}
Efstathios Stamatatos. 2009.
\newblock A survey of modern authorship attribution methods.
\newblock \emph{J. Am. Soc. Inf. Sci. Technol.}, 60(3):538--556.

\bibitem[{Stamatatos et~al.(2018)Stamatatos, Rangel, Tschuggnall, Stein,
  Kestemont, Rosso, and Potthast}]{stamatatos2018overview}
Efstathios Stamatatos, Francisco Rangel, Michael Tschuggnall, Benno Stein, Mike
  Kestemont, Paolo Rosso, and Martin Potthast. 2018.
\newblock Overview of pan 2018.
\newblock In \emph{International Conference of the Cross-Language Evaluation
  Forum for European Languages}, pages 267--285. Springer.

\bibitem[{Tang et~al.(2015)Tang, Qu, Wang, Zhang, Yan, and Mei}]{tang2015line}
Jian Tang, Meng Qu, Mingzhe Wang, Ming Zhang, Jun Yan, and Qiaozhu Mei. 2015.
\newblock Line: Large-scale information network embedding.
\newblock In \emph{Proceedings of the 24th international conference on world
  wide web}, pages 1067--1077. International World Wide Web Conferences
  Steering Committee.

\bibitem[{Thompson(2011)}]{thompson2011radicalization}
Robin Thompson. 2011.
\newblock Radicalization and the use of social media.
\newblock \emph{Journal of strategic security}, 4(4):167--190.

\bibitem[{Tsikerdekis and Zeadally(2014)}]{tsikerdekis2014multiple}
Michail Tsikerdekis and Sherali Zeadally. 2014.
\newblock Multiple account identity deception detection in social media using
  nonverbal behavior.
\newblock \emph{IEEE Transactions on Information Forensics and Security},
  9(8):1311--1321.

\bibitem[{Vaswani et~al.(2017)Vaswani, Shazeer, Parmar, Uszkoreit, Jones,
  Gomez, Kaiser, and Polosukhin}]{vaswani2017attention}
Ashish Vaswani, Noam Shazeer, Niki Parmar, Jakob Uszkoreit, Llion Jones,
  Aidan~N Gomez, {\L}ukasz Kaiser, and Illia Polosukhin. 2017.
\newblock Attention is all you need.
\newblock In \emph{Advances in Neural Information Processing Systems}, pages
  5998--6008.

\bibitem[{de~la Vega and Ng(2018)}]{de2018modeling}
Luis Gerardo~Mojica de~la Vega and Vincent Ng. 2018.
\newblock Modeling trolling in social media conversations.
\newblock In \emph{Proceedings of the Eleventh International Conference on
  Language Resources and Evaluation (LREC-2018)}.

\bibitem[{Volkova et~al.(2015)Volkova, Bachrach, Armstrong, and
  Sharma}]{volkova2015inferring}
Svitlana Volkova, Yoram Bachrach, Michael Armstrong, and Vijay Sharma. 2015.
\newblock Inferring latent user properties from texts published in social
  media.
\newblock In \emph{Twenty-Ninth AAAI Conference on Artificial Intelligence}.

\bibitem[{Vosoughi et~al.(2016)Vosoughi, Vijayaraghavan, and
  Roy}]{Vosoughi:2016:TLT:2911451.2914762}
Soroush Vosoughi, Prashanth Vijayaraghavan, and Deb Roy. 2016.
\newblock Tweet2vec: Learning tweet embeddings using character-level cnn-lstm
  encoder-decoder.
\newblock In \emph{Proceedings of the 39th International ACM SIGIR Conference
  on Research and Development in Information Retrieval}, SIGIR '16, pages
  1041--1044, New York, NY, USA. ACM.

\bibitem[{Waller and Anderson(2019)}]{Waller:2019:GSU:3308558.3313729}
Isaac Waller and Ashton Anderson. 2019.
\newblock Generalists and specialists: Using community embeddings to quantify
  activity diversity in online platforms.
\newblock In \emph{The World Wide Web Conference}, WWW '19, pages 1954--1964,
  New York, NY, USA. ACM.

\bibitem[{Wang et~al.(2014)Wang, Song, Leung, Rosenberg, Wang, Philbin, Chen,
  and Wu}]{wang2014learning}
Jiang Wang, Yang Song, Thomas Leung, Chuck Rosenberg, Jingbin Wang, James
  Philbin, Bo~Chen, and Ying Wu. 2014.
\newblock Learning fine-grained image similarity with deep ranking.
\newblock In \emph{Proceedings of the IEEE Conference on Computer Vision and
  Pattern Recognition}, pages 1386--1393.

\bibitem[{Yang et~al.(2016)Yang, Chang, and Eisenstein}]{yang2016toward}
Yi~Yang, Ming-Wei Chang, and Jacob Eisenstein. 2016.
\newblock Toward socially-infused information extraction: Embedding authors,
  mentions, and entities.
\newblock In \emph{Proceedings of the 2016 Conference on Empirical Methods in
  Natural Language Processing}, pages 1452--1461.

\bibitem[{Zampieri et~al.(2019)Zampieri, Malmasi, Nakov, Rosenthal, Farra, and
  Kumar}]{zampieri2019semeval}
Marcos Zampieri, Shervin Malmasi, Preslav Nakov, Sara Rosenthal, Noura Farra,
  and Ritesh Kumar. 2019.
\newblock Semeval-2019 task 6: Identifying and categorizing offensive language
  in social media (offenseval).
\newblock In \emph{Proceedings of the 13th International Workshop on Semantic
  Evaluation}, pages 75--86.

\end{thebibliography}
